\magnification 1150
\baselineskip 14 pt
\newbox\Ancha
\def\gros#1{{\setbox\Ancha=\hbox{$#1$}
   \kern-.025em\copy\Ancha\kern-\wd\Ancha
   \kern.05em\copy\Ancha\kern-\wd\Ancha
   \kern-.025em\raise.0433em\box\Ancha}}
\font\bigggfnt=cmr10 scaled \magstep 3
 2
\font\bigfnt=cmr10 scaled \magstep 1

\def\Par{\par\vskip 5 pt}
\def\eq#1{ \eqno({#1}) \qquad }
\vglue .3 in
\def\ie{{\it i.\ e.\ }}
\def \ni{\noindent}
\def\brad{|n_1\,j_1\,\epsilon_1\rangle}
\def\brai{ \langle n_2\,j_2\,\epsilon_2|}
\def\bradd{|n_1\,j\, \epsilon\,\rangle}
\def\braid{ \langle n_2\,j\,\epsilon\,|}
\def\brakd{|n_1\,j_1\,\epsilon_1\rangle}
\def\braki{\langle n_2\,j_2\,\epsilon_2|}

\def\ie{{\it i.e.}}
\ni{\bigggfnt  Relativistically extended Blanchard recurrence relation for hydrogenic matrix elements}\par
\vskip 10 pt

\noindent {\bigfnt R.\ P.\ Mart\'{\i}nez-y-Ro\-mero}\footnote{*}{\rm   E-mail: rodolfo@dirac.fciencias.unam.mx} 

\ni {\it Facultad de Ciencias, Universidad Nacional  Aut\'onoma  de M\'exico,\par\ni Apartado Postal 50-542, M\'exico City, Distrito Federal C.\ P.\ 04510.}\par
\vskip 8 pt
\noindent{\bigfnt H.\ N.\ N\'u\~nez-Y\'epez\footnote{\dag}{\rm E-mail: nyhn@xanum.uam.mx}}\par

\noindent{\it Departamento de F\'{\i}sica, Universidad Aut\'onoma Metropolitana-Iztapalapa}\par
\noindent{\it Apartado Postal 55--534, Iztapalapa, Distrito Federal C. P. 09340  M\'exico. }
\vskip 8 pt

\ni {\bigfnt A.\ L.\ Salas-Brito} \footnote{\ddag}{\rm 
E-mail: asb@correo.azc.uam.mx}\par
\noindent{\it Laboratorio de Sistemas Din\'amicos, Departamento de Ciencias B\'asicas, \par 
\noindent Universidad Aut\'onoma Metropolitana-Az\-ca\-pot\-zalco.\par
\noindent  Apar\-tado Pos\-tal 21--726, Co\-yoa\-c\'an, Distrito Federal C. P. 04000  M\'exico. }\Par
\vskip 10 pt

\vskip 20pt
\vskip0.21truein

\centerline{\bigfnt Abstract.}\Par
\vskip0.035truein
\leftskip .3 in \rightskip .3 in \noindent    \Par
\vskip 2pt

General recurrence relations for arbitrary  non-diagonal, radial hydrogenic matrix elements are  derived  in Dirac relativistic quantum mechanics. Our approach is based on a generalization of the second hypervirial method previously employed  in the non-relativistic Schr\"odinger case.    A relativistic version of the Pasternack-Sternheimer relation is thence obtained in the diagonal (\ie\ total angular momentum and parity the same) case,  from such relation an expression for the relativistic virial theorem is deduced.  To contribute to the utility of the relations,  explicit expressions for  the radial matrix elements of functions of the form $r^\lambda$ and  $\beta r^\lambda$  ---where $\beta$ is a Dirac matrix--- are presented.

\vfil

\vskip 10pt
\noindent PACS: 3.65.Ca\par

\eject
\noindent{\bf I. Introduction}\Par	      

 Recurrence relations for matrix elements are  very useful tools in quantum  calculations [1--5]  since the direct computation of such elements is generally very cumbersome.  An interesting example is the  Blanchard's relation which is a useful  recurrence formula for arbitrary  (\ie\  not necessarily diagonal)  non-relativistic matrix elements of the form  $\langle n_1 l_1 | r^\lambda |n_2 l_2 \rangle$,   where the $|n l \rangle$ stand for   
non-relativistic hydrogenic radial energy eigenstates;  according to this relation, once  any three successive matrix elements of powers of the radial coordinate, $r$, are known, any other can be deduced in terms of the previously known ones.  The Blanchard recurrence relation  was derived more than twenty five years ago using a calculation-intensive method [6], which  is not surprising since in general  recurrence relations are rather difficult to obtain. 

Trying to overcome such difficulties, different approaches have been proposed for obtaining general recurrence relations; some of them are based on   algebraic methods,  others use sum rules and  hypervirial theorems [4,7--10].  In particular, a hypervirial result  has been employed to obtain the Blanchard relation in a  more compact way than in the  original deduction [11]. In relativistic quantum mechanics, on the other hand, and despite its physical and its possible chemical interest [7,8,12,13], excepting for the  results reported in [14], there are not as yet general recurrence relations  which could be used for  calculating    matrix elements of  powers of the radial coordinate in terms of   previously known ones.  We should mention though that there are previous efforts in such direction, in which  closed forms for certain relativistic matrix elements have been evaluated [16,17]  and certain, mostly diagonal,  recurrence relations for relativistic and quasirelativistic states have been calculated [3,9,12,17]  .

In this paper we employ  a relativistic calculation inspired  on the hypervirial method [11] to deduce a recurrence relation for the, in general, non-diagonal radial matrix elements of succesive powers of $r^\lambda$  and of $\beta r^\lambda$---where $\beta$ is a 4$\times$4 Dirac matrix [18--20]--- for relativistic  hydrogenic states in the energy basis. The assumptions we use are that the nucleus is point-like and  fixed in space, and that a description using the Dirac equation is valid.  We first study the recurrence relations in the general  case, in which the matrix elements are taken between states with different principal quantum numbers $n_1\neq n_2,$, different total angular momentum quantum numbers  $ j_1\neq j_2$, $m_{j_1}\neq  m_{j_2}$, and, as we use the quantum number $\epsilon\equiv (-1)^{j+l-1/2} $ instead of parity for labelling the hydrogenic eigenstates, where  $\epsilon_1\neq \epsilon_2$. We find that in general the recurrence relations depend on  matrix elements of both powers of $r$ and  of $\beta r$. In practical terms this means that we need two recurrence relations  as the relativistic version of the single-equation  Blanchard relation  [Eqs.\ (6) and (7) of section II].  Given its special interest, we  study in particular the case where the total angular momentum and parity become equal, $j_1=j_2$ and $\epsilon_1=\epsilon_2$, in the two states---not mattering the relative values of the principal quantum number  $n$. We  also address the completely diagonal case where $ n_1=n_2 $, $ j_1 =j_2 $,  and $\epsilon_1=\epsilon_2$. Both of the particular cases mentioned above require special treatment for avoiding possible divisions by zero in the general expressions;  such results  are immediately used to obtain a relativistic version of the Pasternack-Sternheimer rule [21] and to obtain an  expression for the relativistic virial theorem  [9,14,22]. \par

This paper is organized as follows. In section II we review the second hypervirial scheme used for deriving the non-relativistic Blanchard relation. In section III, after obtaining the radial Hamiltonian useful for implementing the hypervirial result in  relativistic quantum mechanics, we proceed to use it to deduce by a  long but direct calculation the relativistic recurrence formulae.  In section IV we study in particular the diagonal case ($j_2=j_1$, $\epsilon_2=\epsilon_1$) to derive the relativistic Pasternak-Sternheimer rule and use it (when $n_1=n_2$) to obtain a version of the relativistic virial theorem. In the Appendix, we obtain explicit expressions for diagonal and non-diagonal matrix elements for any power of $r$ and of $\beta r$ between radial relativistic hydrogenic states. As it becomes evident, such results  are rather cumbersome for relatively large values of the power; for small values, on the other hand, they are better regarded as starting values for the recurrence relations derived in section III of this article. Furthermore, these results can be of utility not only for relativistic atomic and molecular studies but also for evaluating matrix elements of interactions designed to test Lorentz and CPT invariance in hydrogen [14,29].\Par

\noindent{\bf II. The non-relativistic recurrence relation }\Par

Both the Blanchard relation and its predecesor the Kramers selection rule, were originally obtained  employing directly the Schr\"odinger equation  together with appropriate boundary conditions and, at least in the former case, a great deal of computations [1,6]. A much simpler approach  based on a generalized hypervirial result and certain Hamiltonian identities has been developed to  simplify the computations leading to the Blanchard relation [11]. This technique seemed to us an appropriate starting point for deriving  relativistic recurrence formulae. It is with such relativistic extension in mind that we  review in this section the hypervirial method as it is applied in non-relativistic quantum mechanics. In this section, as in most of the paper,  we employ atomic units $\hbar=m=e=1$.

 The idea is to start with the radial Schr\"odinger equation for  a central potential $V(r)$ written in the form

$$  H_k\, |n_k\,l_k\rangle = E_{n_k\, l_k} |n_k\,l_k\rangle, \eq{1}$$

\ni where $ |n_k\,l_k\rangle = \psi_{n_{k}l_k}(r)$ and $E_{n_k\, l_k}$ are  an  energy eigenfunction and its  corresponding energy eigenvalue with  principal and angular momentum quantum numbers,  $n_k$ and  $l_k$, respectively; $k$ is just a label,  and  $H_k$, the non-relativistic radial Hamiltonian, is given by

$$  H_k = - {1\over 2}{d^2\over dr^2} - {1\over r}  {d\over dr}  + {l_k\,(l_k + 1)\over 2r^2} + V(r). \eq{2}$$

 Although we want to calculate the radial matrix elements of terms of the form $r^\lambda$, it is best for our purposes to consider first matrix  elements of  an arbitrary radial function $f(r)$.  With such choice we can readily show [11] that

$$(E_i -E_k) \langle n_i\,l_i\,| f(r)|n_k\,l_k\rangle = \langle n_i\,l_i\,|\Bigl( -{1\over 2} f^{''} - f^{' }{d\over dr}  - {1\over r} f^{'} + {\Delta^-_{ik}\over 2}{f\over r^2}\Bigr)|n_k\,l_k\rangle, \eq{3}$$

\ni where  we use $\Delta^-_{ik} \equiv l_i\,(l_i +1) - l_k\,(l_k +1)$, $E_k\equiv E_{n_k\,l_k}$, and the primes stand for radial derivatives. Please recall that the matrix element of an arbitrary radial function $f(r)$ is  

$$\langle n_i\,l_i\,| f(r)|n_k\,l_k\rangle = \int_0^\infty r^2  \psi^*_{n_{i}l_i}(r)\, f(r)\psi_{n_{k}l_k}(r) dr. \eq{4}$$

\ni To establish   the  result we are after, we apply  the previous result (3) to the  radial function $\xi(r)\equiv H_i f(r) - f(r)H_k$, to find  

$$  \eqalign {
2 ( E_i - E_k)^2\langle n_i l_i| &f(r)|n_k l_k\rangle = \cr  &\langle n_i\,l_i\,|  \bigl( H_i\,(H_if(r) - f(r)H_k)- (H_if(r) - f(r)H_k)H_k  + \cr 
& H_i \,(H_i f(r) - f(r) H_k)- (H_i f(r) - f(r) H_k) H_k \bigr)
|n_k\,l_k\rangle.
}
 \eq{5} $$

\ni  This  is the generalized second hypervirial valid for arbitrary radial potential energy functions, $V(r)$, introduced in   Eq.\  (8) of Ref.\ 11.\par  

The second hypervirial takes a particularly simple form when $f(r)$ is a power of the position, let us say $f(r) = r^{\lambda + 2}$; using this expression for $f(r)$ and  restricting ourselves to the Coulomb potential, $ V(r) = -{Z/ r}$, we obtain [11], after a long ---but much shorter than in [6]--- direct calculation, the Blanchard relation

$$\eqalign{ { \lambda} \left( E_i -  E_k \right)^2 \langle n_i\,l_i\,| r^{\lambda + 2}|n_k\,l_k\rangle& = c_0 \langle n_i\,l_i\,| r^\lambda|n_k\,l_k\rangle  + c_1  \langle n_i\,l_i\,|r^{\lambda -1}|n_k\,l_k\rangle\cr  &+ c_2  \langle n_i\,l_i\,| r^{\lambda-2}|n_k\,l_k\rangle;  }\eq{6}$$

\ni where the hydrogenic energy eigenvalues are $E_a=-Z^2/2n_a^2$, independent of $l$, and   

$$ \eqalign{c_0 &=  Z^2(\lambda +1) {\left[ (l_i - l_k)(l_i + l_k +1)\left({1\over n_i^2} - {1\over n_k^2} \right) +\lambda(\lambda +2)\left(  {1\over n^2_k } + {1\over n^2_i}\right)     \right]}\cr
c_1 &= -2Z\lambda(\lambda + 2)(2\lambda +1)\cr
c_2 &= {1\over 2}(\lambda + 2) \left[ {\lambda}^2 - (l_k - l_i)^2 \right] \left[ (l_k + l_i +1)^2 -\lambda^2 \right]. }\eq{7}  $$

 From this result we can also obtain, as special cases of the Blanchard recurrence relation (6), first the Paster\-nack-Sternheimer selection rule [21]:

$$ \langle n_i\,l_i\,| {Z\over r^2}|n_k\,l_k\rangle  = 0, \eq{8} $$

\ni saying that the matrix element of the potential $1/r^2$ vanishes between  radial states of central potentials when their angular momenta coincide and when the corresponding energy eigenvalues  depend  on the principal quantum number only. Second, in the completely diagonal case (\ie\ $n_i=n_k$, $l_i=l_k$), we can further obtain the non-relativistic quantum virial theorem
[9]

$$  \langle V\rangle = -Z\langle{1\over r}\rangle = 2\langle E\rangle. \eq{9}$$

\noindent As we exhibit in section IV, we can obtain analogous results  using our recurrence relations in relativistic quantum mechanics. \Par

\noindent{\bf III. The relativistic recurrence relations.}\Par

In this section we apply the method  sketched in  section II to the relativistic Dirac case. We clearly need to start with a radial Dirac Hamiltonian analogous to (2). To obtain such Hamiltonian we start with the Dirac Hamiltonian $H_D$ and the corresponding time-independent Dirac equation for  a central potential

$$ H_D = c{\gros \alpha}\cdot\hbox{\bf p} + \beta c^2  + V(r), \quad H_D\Psi({\bf r})=E\Psi({\bf r}); \eq{10}$$

\noindent   where  we are using atomic units, ${\gros\alpha}$ and $\beta$ are the 4$\times$4 Dirac matrices [18--20], which in the Dirac representation are given by

$$ {\gros \alpha}= \pmatrix{ 0&  {\gros\sigma}\cr
                           {\gros\sigma}&   0 }, \qquad \beta=\pmatrix{1& 0\cr
               0& -1}, \eq{11}  $$

\noindent where the 1's and 0's stand respectively, for $2\times2$ unit and zero matrices and the $\gros\sigma$ is the vector composed by the three $2 \times 2$ Pauli matrices ${\gros \sigma}=(\sigma_x, \sigma_y, \sigma_z)$. Please notice that, despite the selection of natural units we shall,  where it aids interpretation, reinsert the appropriate dimensional factors in certain equations. The energy eigenvalues are given explicitly in Eq.\ (63) of section V. The  Hamiltonian $H_D$  is   rotationally invariant, hence the solutions of the Dirac equation (10) can be written in the alternative but entirely equivalent forms [19,23]

$$ 
\Psi(r,\theta,\phi) = {1\over r}\left( \matrix{F_{n j \epsilon}(r){\cal Y}_{jm_z}(\theta, \phi)\cr \cr iG_{n j \epsilon}(r){\cal Y}'_{jm_z}(\theta,\phi)}\right)={1\over r}\left( \matrix{F_{n\kappa}(r){\chi}_{\kappa m_z}(\theta, \phi)\cr \cr iG_{n\kappa}(r){\chi}_{-\kappa m_z}(\theta,\phi)}\right), \eq{12} $$

\noindent where $\chi_{\kappa m_z}$ and $\chi_{-\kappa m_z}$, or ${\cal Y}_{jm}$ and ${\cal Y}'_{jm}$, are  spinor spherical 
harmonics of opposite parity, and $\kappa=-\epsilon(j+1/2)$ is the eigenvalue of the operator $ \Lambda\equiv \beta(1+{\gros\Sigma}\cdot{\bf L})$ which commutes with $H_D$ (where ${\gros \Sigma}\equiv {\gros\sigma}\otimes I=\hbox{diag}({\gros\sigma},{\gros\sigma}) $). The second form in (12) is the preferred in Ref.\ 18. Parity is a good quantum number in the problem because central potentials are invariant under reflections; parity varies as $(-1)^l$ and, according 
to the triangle's rule of addition of momenta,  the orbital angular 
momentum  is given by $l=j\pm {1/2}$. But, instead of working 
directly with parity or with $\kappa$, we prefer  the quantum numbers $j$ and $\epsilon$,  
introduced above, which can be shown also to be

$$ \epsilon =\cases{1 & If $ l=j + {1\over 2},$\cr
\cr
-1 & If $ l= j- {1\over 2}$,}
\eq{13} $$

\noindent thus $l=j+{\epsilon/ 2}$  in all cases.  We also  define $ l'=j 
- {\epsilon/ 2}$; accordingly, the spherical spinor ${\cal Y}_{jm}$ 
depends on $l$ whereas the spherical spinor ${\cal Y}'_{jm}$, which has the 
opposite parity, depends on $l'$. Writing  the solutions in the form (12) completely solves the angular part of the problem. \Par

To construct the radial Hamiltonian, we use the relation 

$$( \gros\alpha\cdot{\bf r})(\gros\alpha\cdot{\bf p})  =  (\gros\Sigma\cdot{\bf r})(\gros\Sigma\cdot{\bf p}) = {\bf r}\cdot{\bf p} + i\gros\Sigma\cdot{\bf L}; \eq{14}$$

\ni   we then use  $ {\bf J}^2 = \left[{\bf L} + (1/2)\gros\Sigma\right]^2 = {\bf L}^2 + \gros\Sigma\cdot{\bf L} + 3/4$ but for expressing the term  $ {\bf L}\cdot\gros\Sigma$, we also need an expression for ${\bf L}^2$ acting on the eigenfunctions (12). Directly from this equation we see that when  ${\bf L}^2 $ is applied to any central potential state, the big component of the state function behaves with the orbital quantum number $l = j+ \epsilon/2$, whereas the small one does so with the orbital quantum number  $l'=j-\epsilon/2$; we have then, 

$$ l(l+1) =j(j+1) + \epsilon(j+{1\over 2}) + {1\over 4},\eq{15}$$
  
\ni for the big component, and 

$$l'(l' + 1) =  j(j+1) - \epsilon(j+{1\over 2}) + {1\over 4},\eq{16} $$

\ni for the small one. The action of ${\bf L}^2$ upon a solution of the form  (12) is therefore always of the form

$$ {\bf L}^2= j(j+1) +\beta \epsilon(j+{1\over 2}) + {1\over 4}, \eq{17}$$

\ni where $\beta$ is the Dirac matrix (11). From this result we obtain the term  $ {\bf L}\cdot\gros\Sigma$  and, substituting it  into $ (\gros\alpha\cdot{\bf p})$, we finally obtain

$$ (\gros\alpha\cdot{\bf p}) =\alpha_r\,[p_r -i\beta{\epsilon\over r} (j+ {1\over 2})], \eq{18} $$

\ni where

$$  \alpha_r \equiv {1\over r} \gros\alpha\cdot{\bf r},  \quad
p_r  = - { i \over r}\left(1 + r{d\over dr}\right). \eq{19}$$

We are now ready to write the relativistic radial Hamiltonian, and the corresponding radial Dirac equation, as

$$ \eqalign{&H_k = c\alpha_r\left[p_r -i\beta{\epsilon_k\over r} \left(j_k+ {1\over 2}\right)\right] +\beta c^2 + V(r), \cr
 &H_k\psi_k(r)=E_k\psi_k(r), }\eq{20}$$

\ni where we introduced the purely radial eigenfunctions

$$ \psi_k(r)\equiv {1\over r}\pmatrix{F_{n_kj_k\epsilon_k}(r)\cr
                                      iG_{n_kj_k\epsilon_k}(r)}\eq{21}      $$

\ni in a $2\times 2$ representation where, $\beta=$ diag$(+1,-1)$, $\alpha_r=\pmatrix{0& -1\cr -1& 0} $, and the radial Dirac equation becomes then [14,19]

$$ \left[\matrix{c^2+\left(V_k(r)-E_k\right)& -c\left(-{\epsilon_k\left(j_k+1/2\right)/ r} +{d/ dr}\right)\cr\cr
          c\left(\epsilon_k\left(j_k+1/2\right)/r +d/dr\right)& -c^2+\left(V_k(r)-E_k\right)}\right] \left[\matrix{F_{n_kj_k\epsilon_k}(r)\cr\cr G_{n_kj_k\epsilon_k}(r)}\right]=0. \eq{22} $$

\noindent Though this explicit representation can  be used for our problem [24,25], it is not really necessary since  all our results are representation independent.\par

The  relativistic recurrence  relation we are after, can be deduced using a  similar reasoning as the used in section II for the non-relativistic case.   Let us first calculate the non-diagonal matrix element  of an arbitrary radial function $f(r)$

$$\eqalign{ (E_2 - E_1) \langle n_2j_2 \epsilon_2|f(r)|n_1 &j_1\epsilon_1\rangle  = \cr &\langle n_2j_2\epsilon_2|H_2f(r)- f(r)H_1|n_1\,j_1\epsilon_1\rangle\cr& = -ic \langle n_2\,j_2 \epsilon_2|\alpha_r\left(f'(r)+ {\Delta^-_{21}\over 2r}\beta f (r)\right)|n_1\,j_1\epsilon_1\rangle,}\eq{23}$$

\ni where from now on the  labelling in the kets stand for the three quantum numbers $n_k$, $j_k$, and $\epsilon_k$, we have defined $  \Delta^-_{21} \equiv \epsilon_2(2j_2 + 1) - \epsilon_1(2j_1 + 1)$, and the matrix elements of radial functions are calculated as

$$\eqalign{ \brai f(r) \brad  &= \int f(r) \left( F^*_2(r)F_1(r)+ G^*_2(r)G_1(r)\right) dr,    \cr
   \brai\beta f(r)\brad & = \int f(r) \left( F^*_2(r)F_1(r) - G^*_2(r)G_1(r) \right)   dr.          }\eq{24}$$ 

\ni where the subscripts stand for the 3 quantum numbers specifying the state.

  We next proceed  to calculate a ``second order iteration'' by substituting $f(r) \rightarrow \xi(r)=H_2f(r)- f(r)H_1 $ in the last expression. Let us calculate first $H_2\xi$ and $ \xi H_1$,

$$\eqalign{&H_2\xi =   -  c^2 \left( {f'(r)\over r} + f''(r) + f'(r){d\over dr} \right)   - c^2 {\Delta^-_{21} \over 2r } \beta \left( f'(r) + f(r){d\over dr} \right) + \cr 
&c^2{\epsilon_2\left( 2j_2 + 1 \right)\over 2 r}\beta \left( f'(r)   +{\Delta^-_{21}\over 2r} \beta f(r)\right)-ic\alpha_r\left(f'(r) + {\Delta^-_{21}\over 2r} \beta f(r) \right)\left( V(r) -\beta c^2 \right),}  \eq{25}$$ 

\ni and 

$$ \eqalign{&\xi H_1=-{c^2\over r}\left( f'(r) -{\Delta^-_{21}\over 2r} \beta f(r) \right)   -c^2\left(f'(r) -{\Delta^-_{21}\over 2r}\beta f(r)  \right){d\over dr}+\cr 
& - c^2{\epsilon_1 \left( 2j_1 + 1 \right) \over 2r}\beta\left( f'(r) -{\Delta^-_{21}\over 2r}\beta f(r) \right)  -i c \alpha_r\left(f'(r) + {\Delta^-_{21}\over 2r} \beta f(r) \right)\left( V(r) +\beta c^2 \right).}\eq{26}$$

\ni  Then, we write down  the difference  of the matrix elements associated with Eqs.\ (25) and (26)

$$\eqalign{ &(E_2-E_1)^2\brai  f(r)   \brad= \cr &\brai-c^2{\Delta^{-}_{21} \over  2r^2} \beta f(r)- c^2 f''(r)-c^2{\Delta^{-}_{21} \over 2r}\beta f'(r) - c^2{\Delta^{-}_{21} \over r}\beta f(r) {d\over dr} + \cr &c^2{\Delta_{21}^{+}\over 2r}\beta  f'(r)  +  c^2\left({\Delta^{-}_{21}\over 2r}\right)^2 f(r) + 2i c^3\alpha_r\beta \left(f'(r) + {\Delta^{-}_{21} \over 2r}\beta f(r)\right)\brad.}\eq{27}$$

\ni where we have defined $ \Delta_{21}^+ \equiv  \epsilon_2 (2j_2 + 1) + \epsilon_1(2j_1 + 1)$. Please notice that here and in what follows we are always assuming $ \Delta^-_{21} \neq 0$.  \par

This last expression (27) is the direct relativistic equivalent of  the generalized second hypervirial [Cf.\ Eq.\ (5) above].   The expression involves the operator $d/dr$, but  here, due to the presence of Dirac matrices in the result, we cannot use the trick employed in the  non relativistic case where we took advantage of the Hamiltonian to simplify the calculation [11]. Instead, let us calculate the following second order iteration for non-diagonal matrix elements

$$  \eqalign{ \brai H_2\xi &+ \xi H_1\brad = (E_2 ^2 - E_1^2)\brai f(r) \brad=\cr
&\brai  \left(-{2c^2f'(r)\over r}  + c^2{\Delta^-_{21} \over 2r^2}\beta f(r) - c^2f''(r) -2c^2f'(r) {d\over dr}  +\right. \cr &\left.{c^2\Delta_{21}^+ \Delta^-_{21} \over 4r^2} f(r) -  2i c\alpha_r \Bigl[f'(r) + {\Delta^-_{21} \over 2r}\beta f(r)\Bigr]V(r) \right) \brad;   } \eq{28}$$

\ni due to the presence of Dirac matrices in our results, we also require  to calculate  non-diagonal matrix elements for expressions involving $\alpha_r f(r) $ and $\beta f(r)$, namely

$$  \eqalign{&H_2\left( -i\alpha_rf(r) \right) =\cr & c\left[ -{f(r)\over r} -f'(r) -f(r) {d\over dr}+{  \epsilon_2\over 2r }\left( 2j_2 +1 \right)\beta f(r)\right] + ic^2\alpha_r\beta f(r) -i\alpha_r V(r) f(r),}\eq{29}$$

\ni and 

$$ \eqalign{ \left( -i\alpha_rf(r) \right)H_1 &=\cr & -cf(r)\left[ {1\over r} \left( 1 + r{d\over dr} \right) + {\epsilon_1\over 2r}\left( 2j_1 + 1 \right)\beta \right]   - ic^2\alpha_r\beta f(r) -i\alpha_r V(r) f(r);}\eq{30} $$

\ni adding up these two last expressions, we get

$$\eqalign{&(E_2 + E_1)\brai -i\alpha_r f(r)\brad = \cr
&\brai -{2cf(r)\over r}  - c f'(r) -2 cf(r) {d\over dr}  + c{\Delta^-_{21} \over 2r}\beta f(r) - 2i\alpha_r V(r) f(r)\brad. }  \eq{31} $$ 

  From the matrix element of $ H_2\left( -i\alpha_r c f(r) \right) - \left( -i\alpha_r cf(r) \right) H_1$, we can obtain 

$$  \eqalign{ (E_2 - E_1)\brai &-i\alpha_r f(r)\brad = \cr
 &\brai  -cf'(r) + c{\Delta_{21}^+ \over 2r} \beta f(r) + 2c^2i\alpha_r\beta  f(r) \brad;}\eq{32} $$

\ni proceeding in a similar way for $ H_2\left(\beta f(r)\right) +\left(\beta f(r)\right)H_1$, we get

$$\eqalign{(E_2 + E_1)\brai \beta f(r)\brad = &
\brai ic\beta\alpha_r f'(r) -i c\alpha_r {\Delta^-_{21} \over 2r} f(r)\cr &+ 2\left[c^2 + \beta V(r)\right]f(r)\brad.} \eq{33}$$

Equations  (23--33) are the basic equations of our problem.  To proceed, we consider, as in the non-relativistic case,  radial functions of the form $f(r) = r^\lambda$  and  insert the explicit expression for the Coulomb potential: $V(r) = -Z/r$. Let us mention though that our results can be generalized to other power of potentials, such as the Lennard-Jones potentials [26]. 

Substituting  $f(r) = r^\lambda$ in (28), it follows 

$$ \eqalign{\left( E_2^2 - E_1^2 \right)&\braki r^\lambda \brakd=\cr
&\braki c^2\left[ {\Delta^+_{21} \Delta^-_{21}\over 4} -\lambda\left( \lambda +1 \right)\right] r^{\lambda -2} + c^2{\Delta^-_{21} \over 2} \beta  r^{\lambda -2} -2c^2\lambda r^{\lambda -1} {d\over dr} +\cr & -2ic\alpha_r\left( \lambda + {\Delta^-_{21}\over 2}\beta \right) r^{\lambda -1} V(r)\brakd  ;}\eq{34} $$

\ni hence,  we can  eliminate the term containing the derivative operator in this last equation, using  $  f(r) = r ^{\lambda -1}$  in Eq.\ (31), to get the result   

$$ \eqalign{\left( E_2^2 - E_1^2 \right)&\braki r^\lambda \brakd=\cr
&\braki  c^2{\Delta^+_{21} \Delta^-_{21}\over 4}r^{\lambda -2} + c^2{\Delta^-_{21} \over 2} \beta \left( 1-\lambda \right) r^{\lambda -2}-i c \alpha_r\beta \Delta^-_{21} r^{\lambda -1} V(r) +\cr & +{\left( E_2 + E_1 \right)}\lambda\left( -ic\alpha_r \right) r^{\lambda -1}\brakd;  }\eq{35} $$

\ni   in this last equation, we use Eq.\ (23) to eliminate the term  with  $ -ic\alpha_r\Delta^-_{21} \beta r^{\lambda -1} $,
to get 

$$ \eqalign{&(E_2^2 - E_1^2)\braki r^\lambda\brakd =\braki c^2\left[{\Delta^-_{21}\Delta_{21}^+ \over 4}   + {\Delta^-_{21}\over 2} (1-\lambda) \beta\right] r^{\lambda -2} +\cr &
2Z\left[ic\alpha_rr^{\lambda -2}  (1-\lambda) - (E_2 - E_1) r^{\lambda -1} \right] - (E_2 + E_1) \lambda ic\alpha_r r^{\lambda -1}\brakd.} \eq{36}$$ 

\ni  Now, from  Eq.\ (32) with  $f(r) = r^{\lambda -1} $ we get 

$$  \eqalign{&(E_2 - E_1)\braki -i\alpha_r r^{\lambda -1}\brakd = \cr
&\braki -c\left( \lambda -1 \right) r^{\lambda -2}  +c{\Delta_{21}^+  \over 2 }\beta r^{\lambda -2}  +2ic^2\alpha_r \beta m r^{\lambda -1}\brakd}\eq{37} $$

\ni and, using $f(r) = r^\lambda $ in Eq. (33) to eliminate the term $ 2ic\alpha_r \beta m r^{\lambda -1} $ from the above equation, we obtain

$$\eqalign{&(E_2 - E_1)\braki -i\alpha_r r^{\lambda -1}\brakd = \cr
&\braki  -c\left( \lambda -1 \right) r^{\lambda -2}  +c{\Delta_{21}^+  \over 2 }\beta r^{\lambda -2}  -{2c \over \lambda}  \left( E_2 + E_1 \right)\beta r^{\lambda}+ {c^2 \over \lambda}\left( -i\alpha_r \right) \Delta^-_{21} r^{\lambda -1}+\cr &
+{4c^3 \over \lambda} r^\lambda - {4cZ \over \lambda} \beta r^{\lambda -1} \brakd ;}\eq{38}$$

\ni which can be written as 

$$\eqalign{& \left[ (E_2 - E_1) - {\Delta^-_{21} c^2\over \lambda} \right]\braki (-i\alpha_r r^{\lambda -1})\brakd =\braki -c(\lambda -1)r^{\lambda -2} +\cr 
& {4c^3\over \lambda} r^\lambda + c{\Delta^+_{21}\over 2} \beta r^{\lambda -2} - {4Zc\over \lambda }\beta r^{\lambda -1} - {2c\over\lambda }(E_2 + E_1) \beta r^{\lambda} \brakd . }\eq{39}$$

We can  also obtain a  new relationship for the matrix elements of  $-i\alpha_r r^{\lambda-1} $, using  Eq.\ (23) with $f(r) = r^\lambda$,  and substitute the result  in Eq.\ (37)  to eliminate the term $ 2i\alpha_r \beta m r^{\lambda -1} $

$$ \eqalign{ &(E_2 - E_1)\braki -i\alpha_r r^{\lambda -1}\brakd = \cr &
\braki  -c\left( \lambda -1 \right) r^{\lambda -2}  + c{\Delta_{21}^+  \over 2 }\beta r^{\lambda -2} + {4c^2\lambda\over \Delta^-_{21}} \left( -i\alpha_r \right) r^{\lambda -1}\cr &
 -{4c \over \Delta^-_{21} }\left( E_2 -E_1 \right) r^\lambda \brakd .} \eq{40} $$

\ni Rearranging terms, we obtain  

$$\eqalign{& \left[ (E_2 - E_1) -{4c^2\lambda\over \Delta^-_{21}}\right]\braki (-i\alpha_rr^{\lambda -1})\brakd=\cr
&\braki-c(\lambda -1 ) r^{\lambda -2}  - {4c\over \Delta^-_{21} }(E_2-E_1)r^\lambda  + c{\Delta^+_{21}\over 2}\beta r^{\lambda -2}\brakd .}\eq{41}$$

The   relation we are looking for  follows  from  this last result and Eq.\ (36).  We use succesively $r^{\lambda -1 }$ and  $r^{\lambda -2}$ from Eq.\ (41) to eliminate the terms  $ 2(E_2 + E_1) \lambda ic\alpha_r r^{\lambda -1} $  and $ 2ic\alpha_rr^{\lambda -2}  (1-\lambda) $ that appear in Eq.\ (36) to  finally get [14]

$$ \eqalign{ c_0 \braki r^\lambda \brakd =\sum_{i=1}^{3} c_i\braki r^{\lambda -i} \brakd + \sum_{i=2}^{3} d_i\braki \beta r^{\lambda -i}\brakd, }  \eq{42}$$

\ni where the numbers $c_i$,  $i=0,\dots 3$ are given by

$$\eqalign{c_0 & = {(E_2^2 - E_1^2)(E_2 - E_1)\Delta^-_{21}\over (E_2 - E_1)\Delta_{21}^- - 4c^2\lambda}, \cr
c_1 & = -{2 Z (E_2 - E_1)^2 \Delta_{21}^-\over (E_2 - E_1)\Delta_{21}^- - 4c^2(\lambda -1)},\cr
c_2 & = c^2{\Delta_{21}^-\Delta_{21}^+\over 4} -c^2\lambda(\lambda -1){(E_1 + E_2)\Delta_{21}^-\over (E_2 - E_1)\Delta_{21}^- -4c^2\lambda},\cr
c_3& = {-2 Zc^2(\lambda -1)(\lambda -2)\Delta_{21}^-  \over (E_2 - E_1) \Delta_{21}^- -4c^2(\lambda -1) },} \eq{43}$$

\ni and the numbers $d_i$, $i=2$ and 3, by

$$ \eqalign{
d_2 & = c^2{\Delta_{21}^-\over 2} \left[     (1-\lambda) + {\lambda (E_2 + E_1)\Delta_{21}^+\over(E_2- E_1) \Delta_{21}^- -4c^2\lambda}\right],\cr
d_3 & = {Z c^2(\lambda -1) \Delta_{21}^- \Delta_{21}^+ \over(E_2 - E_1) \Delta_{21}^- - 4c^2
(\lambda -1)}.}\eq{44} $$

As we may have expected, we need to know six matrix coefficients instead of  only three as in  the non-relativistic case. This is a consequence of the fact that, in the Dirac case, we have to deal with the big and the small components in the state function, doubling in this sense the ``degrees of freedom'' of the system.\par 

It does not seems to be possible to avoid the $\beta$-dependency  in  Eq.\ (44), and thus, taken on its own, Eq.\ (41) does not allow the comp\-utation of $\braki r^\lambda \brakd $ in terms of the $\braki r^{\lambda-a}\brakd$, $a=1, 2, 3$.  The situation is not hopeless though because it is still possible to obtain  another recurrence relation for non-diagonal matrix elements of $\beta r^\lambda $ simply  by eliminating the term $-i\alpha_r r^{\lambda -1}$ between Eqs.\ (39) and (41). In such a way we get

$$  \eqalign{e_0  \braki \beta &r^\lambda \brakd = b_0 \braki r^{\lambda}\brakd + b_2 \braki r^{\lambda-2}\brakd \cr &+ e_1 \braki \beta r^{\lambda-1}\brakd  + e_2 \braki \beta r^{\lambda-2}\brakd,} \eq{45} $$

\noindent where the numbers $b_i$ and $e_i$ $i=1, 2, 3$ are given by

$$ \eqalign{b_0=& 4\lambda\left[(E_2-E_1)^2 -4 c^4 \right], \cr
             b_2=&c^2(1-\lambda)\left[(\Delta_{21}^{-})^2-4\lambda^2\right], \cr
             e_0=&2(E_2+E_1)[(E_2-E_1)\Delta^-_{21}-4c^2\lambda],\cr
             e_1=&4Z[4c^2\lambda-(E_2-E_1)\Delta^-_{21}],\cr
             e_2=& c^2{\Delta_{21}^+\over 2}[(\Delta_{21}^{-})^2-4\lambda^2].} \eq{46} $$

\ni Equations (42) and (45) together are the useful recurrence relations in the relativistic Dirac case.\Par

\noindent{\bf IV. The diagonal case $ \Delta^-_{21}=0$ ($j_2=j_1$, $\epsilon_2=\epsilon_1$).}\Par

In the  results of the last section we always assume $ \Delta^-_{21}\neq 0$, but in order to study the diagonal case we must have $\epsilon_1=\epsilon_2$ and $j_1 =j_2$;  this in turn imply  $\Delta^-_{21}=0$ and  (as always!) $\Delta_{21}^+\equiv\Delta^+\neq 0$. To deal with the diagonal case we start all over again. The equation set for this case is particularly simple, first from Eq.\ (23) we have 

$$ \eqalign{  (E_2 -E_1)\braid f(r) \bradd &= \braid (-ic\alpha_rf'(r)\bradd, \cr  }\eq{47}$$

\ni then we can procced to calculate the second order iteration by substituting, as in the previous section, $f(r) \rightarrow \xi_-=H_2f(r)- f(r)H_1 $ in (47) to obtain

$$\eqalign{(E_2 - E_1)^2\braid f(r) \bradd = \braid  &-c^2f''(r) + c^2{\Delta^+\over  2r} f'(r)\beta +\cr & 2ic^3 \alpha_r \beta f'(r)\bradd ;\cr} \eq{48} $$

\ni and then substitute $ f(r) \rightarrow \xi_+=H_2f(r)+ f(r)H_1 $ again in (47) to get instead

$$\eqalign{(E_2^2 - E_1^2) \braid f(r) \bradd =\braid  & -{2c^2f'(r)\over r} -c^2f''(r) -2c^2f'(r) {d\over dr}\cr &  -2ic\alpha_rf'(r) V(r)\bradd.\cr }\eq{49}$$

\ni The equations  equivalent of Eqs.\ (31--33)  are 
in this case

$$\eqalign{(E_2 + E_1) \braid (-i\alpha_r f(r)) \bradd = \braid &-{2cf(r)\over r}  -cf'(r) -2cf(r) {d\over dr}\cr & -2i\alpha_r V(r) f(r)\bradd, \cr}\eq{50}$$

\ni and

$$\eqalign{(E_2 - E_1) \braid (-i\alpha_r f(r) )\bradd = \braid  & -cf'(r) + {c\Delta^+\over 2r} \beta f(r) +\cr &2i\alpha_r\beta c^2 f(r) \bradd.\cr }\eq{51}$$

\ni We also have,  for  the matrix elements of $\beta f(r)$,  

$$\eqalign{(E_2 + E_1)\braid \beta f(r) \bradd = \braid  & -ic\alpha_r\beta f'(r) +\cr & 2\left[c^2 + \beta V(r)\right] f(r) \bradd. }  \eq{52} $$

\ni These expressions are the basic equations for the case $ \Delta^-_{21}=0$.\par

We can  now obtain a recurrence relation valid in the diagonal case. First, let us use $f(r) = r^\lambda$ in Eq.\ (48) to  get

$$\eqalign{(E_2-E_1)^2\braid r^\lambda \bradd  = &\lambda  \braid -c^2(\lambda -1)r^{\lambda -2} + c^2{\Delta^+\over 2} \beta r^{\lambda-2} +\cr &2ic^3\alpha_r\beta  r^{\lambda -1}\bradd.}\eq{53} $$

\ni Evaluating now equation (52) with $f(r)= r^\lambda $, we obtain

$$ \eqalign{(E_2 + E_1) \braid\beta r^\lambda \bradd =& \braid-ic\alpha_r\beta \lambda r^{\lambda -1} \cr &+ 2\left( c^2 - {Z\beta\over r}\right)r^\lambda\bradd, } \eq{54}$$

\ni and eliminating the $i c\alpha_r \beta\lambda r^{\lambda-1}$ between Eqs.\ (53) and (54),  we finally get 

$$ \eqalign{ &\left[ (E_2 - E_1)^2 - 4c^4 \right]  \braid r^\lambda \bradd = \lambda c^2 {\Delta_{21}^+\over 2}\braid\beta r^{\lambda-2}\bradd \cr & -4Zc^2\braid\beta r^{\lambda-1}\bradd-2c^2(E_2+E_1)\braid\beta r^{\lambda}\bradd\cr& -c^2\lambda(\lambda-1) \braid r^{\lambda-2}\bradd.} \eq{55} $$

\ni This is the only recurrence relation we get in the diagonal case. To ``close'' the  relation we can use the diagonal recurrence relations given in  [9].

The special case when $\lambda =0$  is of particular interest

$$ \eqalign{\left[ (E_2 - E_1)^2 - 4c^4 \right]\delta_{n_1 n_2}&=-4Zc^2\braid{\beta\over r}\bradd\cr
&-2c^2 (E_2 + E_1) \braid\beta\bradd.}\eq{56} $$

\ni This expression could be considered as a relativistic generalization of the Paster\-nak-Sternheimer  rule of non relativistic quantum mechanics (Equation (8) of section II) [21], which says that the expectation   value between hydrogenic states of the $1/r^2$ potential,  vanishes when  the orbital angular momenta of the  states $1$ and $2$ coincide, \ie\ when $l_1=l_2$.  In the relativistic case the expectation value of the  $\beta/r$ potential (which could be regarded as  the square root of $1/r^2$ including {\sl both} signs),  does {\bf not} necessarily vanish even when the total angular momenta of the two states coincide: \ie\ it does not vanish when $j_1=j_2$.  Again, this agrees with the  fact that the non-relativistic Pasternack-Sternheimer rule is applicable to eigenfunctions of potentials whose energy eigenvalues depend only on the principal quantum number---which is not the case for the hydrogen atom in Dirac relativistic quantum mechanics [14].\par

Moreover, two special cases  are immediately deduced from this last expression (56):

\ni 1) The first case, when  $n_1\neq n_2$, is

$$  \braid {Z\beta\over r} \bradd = - {1\over 2}(E_2 + E_1) \braid \beta\bradd.\eq{57}$$

\ni 2) The other case follows when $n_1 =n_2$

$$c^2=-\left< \beta V(r)\right> +  E\,  \left<  \beta \right> =Z\left<{\beta\over r}\right>
 + E\,  \left<  \beta \right> ,\eq{58} $$

\ni which is  the relativistic virial theorem [22]; from the  relation  $c^2<\beta> =E$ [9], we can also
put it in the alternative form

$$ E^2=c^2\, \left< \beta V(r)\right> +c^4 =-c^2\,Z \left<  {\beta\over r}\right> +c^4.\eq{59} $$
\Par

\noindent{\bf V. The values of $ <r^\lambda>$ and  $ < \beta r^\lambda >$.}\Par

The recurrence relations found above, involve in principle simple expressions  (since they  involve only matrix elements of  Dirac hydrogenic states) that can be burdensome to handle. Given such situation, we have also calculated explicit formulas that are needed to  evaluate  the  diagonal and the non-diagonal matrix elements of interest. The expressions  are related to the hypergeometric function and can be deduced  from the two differential equations that follow  from the Hamiltonian  (20), as it is shown in the Appendix. In particular,  from Eq. (A.15)  we  calculate $ <r^\lambda>$  and $<\beta r^\lambda>$. We quote the results here and refer to the Appendix for the details. 

$$ <r^\lambda> = {mc^2|C|^2\over (2k)^{\lambda+1} 2^{s-1}}\left[I^{2s}_{nn}(\lambda) u^2 + I^{2s}_{n-1 n-1}(\lambda) v^2 + E\,uvI^{2s}_{nn-1}(\lambda) \right],\eq{60a}$$

\ni and

$$ <\beta r^\lambda> = {E|C|^2\over (2k)^{\lambda+1} 2^{s-1}}\left[I^{2s}_{nn}(\lambda) u^2 + I^{2s}_{n-1 n-1}(\lambda) v^2 + mc^2\,uvI^{2s}_{nn-1}(\lambda) \right];\eq{60b}$$

\ni in these expressions  $n=0, \, 1, \,2, \cdots$,  and  [23,27]

$$\eqalign{k\equiv {1\over \hbar c}& \sqrt{m^2 c^4  -E^2}, \quad \zeta\equiv {Ze^2\over \hbar c}= Z\alpha_F,\quad \tau_j\equiv \epsilon(j+{1\over 2}), \cr \cr &\nu \equiv \sqrt{mc^2-E\over mc^2+ E}, \quad s\equiv\sqrt{\tau_j^2 -\zeta^2},}
\eq{61} $$

\ni where $\alpha_F\simeq 1/137$ is the fine structure constant and the $I^{2s}_{nm}(\lambda)$ symbols are defined in equation (A.15) of the Appendix.  The numbers $u$ and $v$ are {\sl constants} such that 

$$u=( \tau_j + s + n - \zeta\nu^{-1}) ^{1/2},\quad v= (n + 2s)( \tau_j + s + n - \zeta\nu^{-1}) ^{-1/2};\eq{62}   $$

\ni in the Appendix we give a simple proof of this result.
 Notice that in this section  we have explicitly written  $\hbar$, $e$, and $c$ in our results. 
 Finally, to obtain $C$, we  use  relations (61) to get $  (\tau_j + s +n -\zeta\nu^{-1})^{-1} =(n + s -\tau_j-\zeta\nu^{-1})/ n(n + 2s)$; we need  also  $ (n + s) = \zeta E/ \sqrt{m^2 c^4 -E^2}$, which is obtained from the expression for the energy eigenvalues of the Dirac hydrogen atom:

$$E=mc^2  \left( 1+{Z^2\alpha_F^2\over \left(n-j-1/2+\sqrt{(j+1/2)^2-Z^2\alpha_F^2}\right)^2}\right)^{-1/2};                       \eq{63}  $$

\ni elementary algebra gives then the result

$$ |C|= {\hbar \,2^{s-{1}}\over Z\alpha c^2} \sqrt{n!\, k\over 2 m^3}\left[   \Gamma(n + 2s + 1)\right]^{-1/2}. \eq{64}$$

\ni where  we have  written  explicitly the dimensional factors. \Par
\vfill
\eject
\noindent{\bf Acknowledgements.}\Par

\noindent This work has  been partially supported by CONACyT. It is a pleasure to thank C.\ Cisneros for all the collaboration,  and V.\ M.\ Shabaev for making us aware of Ref.\ 15. ALSB and HNNY acknowledge the help of F.\ C.\ Minina,  B.\ Caro,  M.\ X'Sac, M.\ Osita, Ch.\ Dochi,  F.\ C.\ Bonito, G.\ Abdul,  C.\ Sabi,   C.\ F.\ Quimo,   S.\ Mahui, R Sammi, M.\ Mati, U.\ Becu, Q.\ Chiornaya, E.\ Hera and M.\ Sieriy. Last but not least, this paper is dedicated to the memory of our beloved friends Q.\ Motita, B.\ Kuro, M.\ Mina, Ch Cori, C.\ Ch.\ Ujaya, Ch.\ Mec, F.\ Cucho, R.\ Micifuz and U.\ Kim.\Par

\vskip 12 pt
\noindent{\bf Appendix. Explicit expressions for relativistic matrix elements of $r^\lambda$ and $\beta r^\lambda$} \Par

It is possible to obtain explicit expressions for the diagonal and non diagonal matrix elements   in the case $V(r)=-Z/r.$ The purpose of this appendix is to give the basic relation that is needed for such evaluation. As we heavily draw from results previously obtained, in this section we use the notation  of Ref.\ 23, in particular $\hbar=c=e=1$, though we sometimes write all the dimensional constants.  

We are interested in the bound states of the problem, so the quantity $k\equiv\sqrt{m^2 - E^2}$ is positive.  We can write the differential equations for the radial part of any central problem in terms of the dimensionless variable $\rho \equiv kr$  [23,25] and the symbols defined in (60)

$$\eqalign{ \left(-{d\over d\rho} + {\tau_j\over \rho}\right)G(\rho)& = \left( -\nu 
+{\zeta\over \rho}\right)F(\rho),\cr
 \left(+{d\over d\rho} + 
{\tau_j\over\rho}\right)F(\rho) &= \left( \nu^{-1} + 
{\zeta\over\rho}\right)G(\rho);}\eq{A1} $$ 

\ni where we look for solutions of the form

$$
F(\rho)  = \sqrt{m + E}\,\left[\psi_{-}(\rho) + \psi_{+}(\rho)\right], \eq{A2} $$

$$
G(\rho)  = \sqrt{m - E}\,\left[\psi_{-}(\rho) - \psi_{+}(\rho) \right]. \eq{A3}
$$

The solution to these coupled differential equations can be written in terms of  the Laguerre polynomials of non-integer index [25,27,28]

$$ \eqalign{
    \psi_+(\rho)=&a\rho^s \exp(-\rho) {\cal L}^{2s}_{n -1}(2\rho), \cr
    \psi_-(\rho)=&b\rho^s \exp(-\rho) {\cal L}^{2s}_{n}(2\rho),  }
     \eq{A4}$$

\ni where the Laguerre polynomials $ {\cal L}_n^{\alpha}(\rho)$  are related to both  the hy\-per\-geo\-metric fun\-ction, $ {}_1F_1(-n,\alpha+ 1; \rho)$, and the Sonine polynomials,  $ T_\alpha^{\,(n)}(\rho)$ [28], through the relation

$$ {\cal L}_n^{\alpha}(\rho)={\Gamma(\alpha+n+1)\over n! \Gamma(\alpha + 1)} {}_1F_1(-n;\alpha+1;\rho) = (-1)^n\Gamma(\alpha +n+1)\,T_\alpha^{\,(n)}(\rho),\eq{A5} $$

\ni and $a$ and $b$ are  constants. Substitution of these results in Eq.\ (A1) gives the condition

$$ \eqalign{a(\tau_j + s - \zeta\nu^{-1} + n) + b(n + 2s) &=0,\cr
b( \tau_j - s + \zeta\nu^{-1} -n) -an &=0.}\eq{A6} $$

\ni Solving these last  two equations give us  a relationship between $n$ and $\nu$. From Eq.\ (45)  we see that we can solve for the energy $E$ and obtain the relativistic energy spectrum (63), provided we first introduce the principal quantum number $ N\equiv j + 1/2 +n$. To proceed further, we  take 

$$ b=-{a( \tau_j + s +n -  \zeta\nu^{-1})/ (n + 2s)}, \eq{A7} $$ 

\ni and write the result in a symmetrized form:

$$ \eqalign{F(\rho) &=\sqrt{mc^2 + E}\,C\rho^se^{-\rho}\left[ u \,{\cal L}_n^{2s} (2\rho) +v\,{\cal L}_{n-1} ^{2s}(2\rho) \right],\cr
G(\rho) &=-\sqrt{mc^2 - E}\,C\rho^se^{-\rho}\left[ u\,{\cal L}_{n}^{2s} (2\rho) -v\,{\cal L}_{n-1} ^{2s}(2\rho) \right],}\eq{A8} $$

\ni where 

$$u =( \tau_j + s + n - \zeta\nu^{-1}) ^{1/2},\quad v= (n + 2s)( \tau_j + s + n - \zeta\nu^{-1}) ^{-1/2},\eq{A9}   $$

\ni  $C$ is a normalization constant that  can be obtained from

$$ \int_0^\infty e^{-x} x^\alpha {\cal L}^\alpha_n(x){\cal L}^\alpha_m(x) = \delta_{mn} {\Gamma(n + \alpha +1)\over n!}; \eq{A10}$$

\ni after some work we obtain 

$$ |C|= {\hbar \,2^{s-{1}}\over Z\alpha c^2} \sqrt{n!\, k\over 2 m^3}\left[   \Gamma(n + 2s + 1)\right]^{-1/2}. \eq{A11}$$

We can also calculate the expectation values for diagonal and non diagonal  matrix elements. For  diagonal, arbitrary power matrix elements of the form $<r^\lambda>$ and $<\beta r^\lambda>$, we need to calculate the expression

$$ I^\alpha_{nm} (\lambda)= \int_0^\infty e^{-x} x^{\alpha + \lambda} {\cal L}_n^\alpha (x)\, {\cal L}^\alpha_m (x) dx.\eq{A12}$$

\ni This expression converges for $Re(\alpha + \lambda +1)>0$, and is  zero if $\lambda$ is an integer such that $m-n>\lambda\geq 0,$ where without loss of generality, we assume that $m>n$.  From Rodrigues formula and $ (d^m/dx^m) x^{k + \lambda}= (-1)^m [-k -\lambda]_mx^{k + \lambda -m}$, where $[n]$, $n$ an integer, is a Pochhammer symbol [28],  we find, after a $m$-times partial integration, 

$$  I^\alpha_{nm} (\lambda)= {1\over m!}\sum^n_{k=0}(-1)^k {\Gamma(n+\alpha+1)\Gamma(\alpha + k + \lambda +1 )[-k-\lambda]_m\over k!\,(n-k)!\,\Gamma(\alpha + k +1)}.\eq{A13} $$

\ni We use now the identity $ [-k-\lambda]_m = [-k-\lambda]_{k}[-\lambda]_{m-k} , $ change the order of summation $k\to n-k$ and use the identities

$$\eqalign{  [-\lambda]_{m-n+k}&= [-\lambda]_{m-n}[-\lambda+m-n]_k,\cr \Gamma(n + \alpha +1)&=(-1)^k \,\Gamma(\alpha + n -k +1) \, [-\alpha -n]_k,\cr
\Gamma(\alpha + \lambda + + n +1 ) &= (-1)^k\Gamma(\alpha + \lambda +n -k +1)[-\alpha-\lambda-n]_k,\cr
[k-n-\lambda]_{n-k} &= (-1)^n {\Gamma(\lambda + n +1)\over \Gamma(\lambda+1)}\, {1\over [-\lambda-n]_k},}\eq{A14}$$

\ni to obtain that $$\eqalign{& I^\alpha_{nm} (\lambda)=\cr&[-\lambda]_{m-n} {\Gamma(\alpha + \lambda + n +1)\Gamma(\lambda+n +1)\over m!\,n!\,\Gamma(\lambda+1)}\,\cr &{}_3F_2(-\alpha-n,-\lambda+m-n,-n;-\lambda-n,-\alpha-\lambda-n;1). }\eq{A15}$$

We consider two cases for the general  matrix elements $ \brai r^\lambda\brad$  and $  \brai \beta r^\lambda \brad$;   the first one, when  $ k_1=k_2$, where we  need to evaluate 

$$K_{nn}^{s_1s_2}(\lambda) = \int_0^\infty x^{s_1 + s_2 +\lambda} e^{-x}{\cal L}_n^{(2s_1)}(x) {\cal L}_m^{(2s_2)} (x) \, dr,\eq{A16}  $$

\ni and the second one, when $ k_1\neq k_2$, where we  need 

$$ K_{nm}^{s_1s_2}(\lambda) = \int_0^\infty r^{s_1 + s_2 +\lambda} e^{-(k_1 + k_2)r}{\cal L}_n^{(2s_1)}(2k_1r) {\cal L}_m^{(2s_2)} (2k_2r) \, dr.\eq{A17}$$

\ni In the first case, we see  that integral  (A16)  is convergent  if  $Re(s_1 + s_2 + \lambda + 1)>0,$  and   vanishes when $ s_1 -s_2 +\lambda $ is an integer such that $ m-n > s_1-s_2 +\lambda\geq 0$. Using a similar reasoning as in the diagonal case, we get

$$ \eqalign{& K^{s_1s_2}_{nm} (\lambda)=\cr&[-\lambda + s_2 -s_1]_{m-n} {\Gamma(s_1 + s_2 + + \lambda + n +1)\Gamma(\lambda+s_1  - s_2 +n +1)\over m!\,n!\,\Gamma(\lambda+s_1 - s_2+ 1)}\,\cr &{}_3F_2(-2s_1-n,-\lambda+s_2 -s_1 +m-n,-n;-\lambda + s_2 -s_1 -n,-\lambda-s_1-s_2-n;1). } \eq{A18}$$

In the second case, the  integral converges for $Re(s_1 + s_2 + \lambda +1)>0 ,$ and is not zero provided $k_1\neq k_2.$ A straightforward calculation by parts shows that

$$\eqalign{   K_{nm}^{s_1s_2}(\lambda)& = \sum_{j=0}^n\sum_{i=0} ^m {(-1)^j(k_2 -k_1)^{m-i}(k_1 +k_2)^{(i-m-s_1-s_2-\lambda-1)}\over i!\,j!\,(m-i)!\,(n-j)!} \cr & {\Gamma  (n+2s_1 +1)[s_2-s_1-\lambda-j]_i\over \Gamma(2s_1 +j +1)}  \Gamma(m+s_2+s_1+\lambda-i +1),}\eq{A19}$$ 

\ni where $k_1\neq k_2.$  Although less practical than the other expressions found here, we still can rewrite Eq. (A19) in a different form using the well known identities
[28]

$$ \eqalign{&[p]_{m-i} = (-1)^{m-i} {\Gamma(-p +1)\over [-p -m +1]_i \,\Gamma(-p -m + 1)},\cr
&\Gamma(-p -m +i +1) =[-p-m +1]_i\Gamma(-p-m+ 1),\cr
&(m-i)!\, = {m!\,(-1)^i \over [-m]_i},\cr
& \Gamma(p + 1) = [p-m+1]_m\Gamma(p-m+1). }\eq{A20}$$ 

\ni for any number $p$  and $m$ and $i$ integers. After some algebra, we finally get

$$\eqalign{&K_{nm}^{s_1s_2}(\lambda)\cr& = {(-1)^m\Gamma(n + 2s_1 +1)\Gamma(s_1 + s_2 +\lambda +1)\over  m!\, (k_2 + k_1)^{s_1 + s_2 + \lambda +1}   }\sum_{j=0}^n {(-1)^j\,[s_1-s_2 +\lambda +j -m +1]_m\over j!\,(n-j)!\,\Gamma(2s_1 +j +1) }\cr
&\times{}_2F_1(-m, s_1 + s_2 + \lambda +1;s_1-s_2 + \lambda +j -m +1; {k_2-k_1\over k_2 +k_1}). }\eq{A21}  $$ 

\vfill
\eject

\noindent{\bf References}\Par

\noindent 1. H.\ A.\ Kramers,  {\it  Quantum mechanics}, (North Holland,  Amsterdam, 1957).\par

\noindent 2.  P.\ W.\ Atkins, {\it Molecular Quantum Mechanics}, (Oxford, Clarendon, 1970). \par

\ni 3. J.\ Kobus, J.\ Karkwowski, and W.\ Jask\'olski, {\it J. Phys. A: Math. Gen. } {\bf 20}, 3347, (1987). \par

\ni 4. B.\ Moreno,  A.\ L\'opez-Pi\~neiro and R.\ H.\ Tipping, {\it J.\ Phys.\ A: Math.\ Gen.}, {\bf 24}, 385, (1991) .\par

\ni 5. H.\ N.\ N\'u\~nez-Y\'epez, J.\ L\'opez-Bonilla, D.\ Navarrete, and A.\ L.\ Salas-Brito, {\it Int.\ J.\ Quantum Chem.} {\bf 62}, 177, (1997).\par 

\ni 6. P.\ Blanchard, {\it  J. Phys. B: At. Mol. Phys.}, {\bf 7},  993, (1974). \par

\ni 7.  F.\ M.\ Fern\'andez and E.\ A.\ Castro, {\it Hypervirial Theorems}, (Springer, New York,  1987).\par

\ni 8. F.\ M.\ Fern\'andez and E.\ A.\ Castro, {\it Algebraic Methods in Quantum Chemistry and Physics}, (CRC, Boca Raton, 1996).\par

\ni 9. O.\ L.\ de Lange and R.\ E.\ Raab, {\it Operator Methods in Quantum Mechanics},  (Clarendon, Oxford, 1991).\par 

\ni 10. J.\ Morales, J.\ J.\ Pe\~na, P.\ Portillo, G.\ Ovando, and  V.\ Gaftoi,  {\it Int.\ J.\ of Quantum Chem.}, {\bf 65}, 205, (1997).\par

\ni 11.  H.\ N.\ N\'u\~nez-Y\'epez, J.\  L\'opez-Bonilla, and A.\ L.\ Salas-Brito, {\it  J.\ Phys.\ B: At.\ Mol.\ Opt.}, {\bf 28}, L525, (1995).\par

\ni 12. M.\ K.\ F.\ Wong  and H-Y.\ Yeh,  {\it Phys.\ Rev.\ A},  {\bf 27}, 2300, (1983).\par

\ni 13. I. V.\ Dobrovolska and R.\ S.\ Tutik, {\it Phys. Lett. A}, {\bf 260}, 10, (1999).\par

\ni 14. R.\ P.\ Mart\'{\i}nez-y-Romero, H.\ N.\ N\'u\~nez-Y\'epez, and A.\ L.\ Salas-Brito, {\it J.\ Phys.\ B: At.\ Mol.\ Opt.\ Phys.},  {\bf 33}, L367, (2000).\par

\ni 15. N.\ Bessis, G.\ Bessis, and D.\ Roux,  {\it Phys.\ Rev.\ A}, {\bf 32}, 2044, (1985).\par

\ni 16. V M Shabaev J.\ Phys.\ B: At.\ Mol.\ Opt.\ Phys.\ {\bf 24}, 4479 (1991).\par

\ni 17. M.\ K.\ F.\ Wong  and H-Y.\ Yeh,  {\it Phys.\ Rev.\ A},  {\bf 27}, 2305, (1983).\par

\ni 18. J.\ D.\ Bjorken and S.\ D.\ Drell,  {\it Relativistic Quantum Mechanics}, (Mac Graw-Hill, New York, 1964). \par

\ni 19.  I.\ P.\ Grant in G.\ W.\ F.\ Drake Editor, {\it Atomic, Molecular and Optical Physics Handbook}, (American Institute of Physics, Woodbury, 1996) Ch.\ 32.

\ni 20. R.\ E.\ Moss, {\it Advanced Molecular Quantum Mechanics}, (London, Chapman and Hall, 1972).\par

\ni 21. S.\ Pasternack, R.\ M.\ Sternheimer, {\it  J.\ Math.\ Phys.},  {\bf  3},  1280, (1962).\par

\ni 22.  Y. S. Kim, {\it Phys. Rev.} {\bf 154}, 17, (1967). \par

\ni 23. R.\ P.\ Mart\'{\i}nez-y-Romero, A.\ L.\ Salas-Brito and J.\ Salda\~na-Vega, {\it J. Math. Phys.},  {\bf 40}, 2324, (1999);

\ni 24. F.\ Constantinescu  and E.\ Magyari, {\it Problems in Quantum Mechanics}, (Oxford, Pergamon, 1971).\par

\ni 25.  R.\ P.\ Mart\'{\i}nez-y-Romero, A.\ L.\ Salas-Brito and J.\ Salda\~na-Vega, {\it J.\ Phys. A: Math. Gen.\ } {\bf 31}  L157 (1998).\par

\ni 26. U.\ Fano  and L.\ Fano,  {\it Physics of atoms and molecules}, (University of Chicago, Chicago, 1972).\par

\ni 27. L.\ Davies Jr.\ {\it Phys. Rev.}, {\bf  56}, 186, (1939).  \par

\ni  28. W.\ Magnus and  F.\ Oberhettinger, {\it Formulas and Theorems for the Special Functions of Mathematical Physics},  (Chelsea, New York, 1949).\par

\ni 29. R.\ Bluhm, V.\ A.\ Kostelecky, N.\ Rusell  Preprint hep-ph/0003223 (2000).\par

\vfill
\eject
\bye